# Three-dimensional representation of the many-body quantum state

Peter Holland
University of Oxford
peter.holland@gtc.ox.ac.uk

**Abstract**

Using the trajectory conception of state we give a simple demonstration that the quantum state of a many-body system may be expressed as a set of states in three-dimensional space, one associated with each particle. It follows that the many-body wavefunction may be derived from a set of waves in 3-space. Entanglement is represented in the trajectory picture by the mutual dependence of the 3-states on the trajectory labels.

**1 Motivation for the spatial trajectory conception of the quantum state**

A curious dichotomy between theory and practice pervades the history of quantum mechanics. On the one hand, the theory is supposed to be about 'measurements', procedures whose outcomes are the eigenvalues of self-adjoint Hilbert space operators that represent the observables 'measured'. The role of the 'state' of a physical system, a vector $\psi(x)$ (in the position representation) in the Hilbert space, is to encode the probabilities of the empirical outcomes. The $\psi$ conception of state has been adopted almost universally since the advent of quantum theory. Both the formalism and the debates over the theory's meaning are routinely couched in its terms, including by those who seek to discern causal mechanisms underlying the statistical $\psi$ calculus.

On the other hand, in real laboratories rather than in theoreticians' heads measurements are about the determination of position – of a meter pointer, of a symbol in a printout, of an oscilloscope track,... The statistical regularities predicted by the theory are tested, in the end, by sequences of individual position-experiments (amplified to the macroscopic level). *Empirical physical assertions about a quantum system are either about or are inferences from the measurement of position*. When we 'measure spin' we infer that quantity from the discrete spatial domains impacted by a beam of identically prepared systems on a detecting screen, the cumulative density of the successive impacts indicating the probability distribution. For all the talk of operators, Hilbert space and entanglement, we have to map our abstract multidimensional theoretical analysis into assertions about the (likely) locations of moving objects in three-dimensional physical space, that is, in the first instance, into statements about three-dimensional *trajectories*.

The following question therefore presents itself: if our direct connection with the 'quantum world' is through the time-varying positions of objects in physical 3-space, which objects may legitimately be regarded as part of an ecumenical quantum description even if they comprise macroscopic components, why is the theory not formulated directly in these terms, that is, why is the quantum state not a time-dependent position variable rather than merely a time-dependent encoder of the statistics of position? To couch the theory directly in terms of experimental outcomes would, after all, chime with the instrumentalist views that have dominated most quantal discourse.

Of course, these are contentious issues. But it turns out that the basic problem that emerges from these considerations, that of representing the quantum state using position as the state variable, has a simple and apparently uncontentious solution [1] (for a recent account and further references see [2] and for a discussion setting the theory in a wider conceptual and historical context see [3]). In fact, the model we propose accounts for more than just empirical variables; it provides an alternative conception of the quantum state in general processes, measurements or otherwise. Moreover, the two state-pictures, the wavefunction and the trajectory, are not in conflict; they stand in a harmonious complementary relation of codetermination. The wavefunction formulation describes temporal changes in the system's state at each space point (analogous to the Eulerian picture in fluid mechanics) and the trajectory formulation describes the transport of the system's state across space (analogous to the fluidical Lagrangian picture). In particular, the paths are conveyors of constant parcels of probability. This extension of the notion of state raises interesting questions about how quantum processes may be comprehended but our objective here is more modest: to highlight that, according to the alternative trajectory formulation, the quantum state of an *n*-body system may be expressed as a set of *n* 3-dimensional states, one associated with each of the *n* particles (this formulation is a special case of the trajectory theory developed for a generalized Riemannian configuration space [4]). Merging aspects of both conceptions of state implies a corresponding three-dimensional decomposition of the $\psi$ version, a construction that has hitherto proved elusive. Entanglement is represented in the trajectory picture by the mutual dependence of the 3-states on the trajectory labels. We do not go further here into other potential roles for the trajectories, such as supporting the flow of matter or enabling causal representations of microprocesses.

## 2 Transformation of Schrödinger's equation into the trajectory picture

A straightforward way to obtain the trajectory theory of the quantum state is to transform the independent variables $x_i$ in the Schrödinger equation for a particle of mass *m* in a potential *V* and wavefunction $\psi(x,t)$,

$$i\hbar \frac{\partial \psi}{\partial t} = -\frac{\hbar^2}{2m}\frac{\partial^2 \psi}{\partial x_i \partial x_i} + V\psi, \qquad (2.1)$$

into dependent variables $x_i = q_i(a,t)$, $i = 1,2,3$, defined by the integral curves of the velocity vector field: $\partial q_i/\partial t = m^{-1} \partial S/\partial x_i\big|_{x=q(a,t)}$ where *S* is the phase ($\psi = \sqrt{\rho}e^{iS/\hbar}$) and $a_i = q_{0i}$ is the initial position. The initial condition $\psi_0(x)$ is transformed similarly. The single-valuedness of the velocity field implies that the trajectories are uniquely labelled by $a_i$, whose variation generates a differentiable congruence of spacetime trajectories $q_i(a,t)$. The wave equation may be recast as a self-contained dynamical equation describing the evolution of the displacement vector $q_i$ as a function of $a_i$ and *t*, supplemented by the appropriate initial conditions corresponding to the initial condition $\psi_0$. The quantum state is then represented by the non-denumerable set of trajectories $q_i(a)$ occupying the space where $\psi(x)$ is finite together with initial conditions on their density and velocity.

This transformation has been examined in detail elsewhere [1,2] so we shall just state the results. The Schrödinger equation for $q_i$ has first- and second-order versions (in time). The first-order form is the integro-differential equation

$$m\frac{\partial q_i}{\partial t}\frac{\partial q_i}{\partial a_k} = m\dot{q}_{0k} + \frac{\partial}{\partial a_k}\int_0^t \left[\tfrac{1}{2}m\frac{\partial q_i}{\partial t}\frac{\partial q_i}{\partial t} - V(q(a,t)) - V_Q(q(a,t))\right]dt. \quad (2.2)$$

This is less useful computationally but important in the formal structure of the theory. Differentiating (2.2) with respect to $t$ yields the equivalent local second-order form, a version of Newton's second law:

$$m\frac{\partial^2 q_i(a,t)}{\partial t^2} = -\frac{\partial}{\partial q_i(a,t)}\left[V(q(a,t)) + V_Q(q(a,t))\right]. \quad (2.3)$$

In these expressions $i,j,k,\ldots = 1,2,3$. The derivatives with respect to $q_i$ are shorthand for derivatives with respect to $a_i$ via the formula

$$\frac{\partial}{\partial q_i} = J^{-1}J_{ij}\frac{\partial}{\partial a_j} \quad (2.4)$$

where

$$J = \det(\partial q/\partial a) = \frac{1}{3!}\varepsilon_{ijk}\varepsilon_{lmn}\frac{\partial q_i}{\partial a_l}\frac{\partial q_j}{\partial a_m}\frac{\partial q_k}{\partial a_n}, \quad 0 < J < \infty, \quad (2.5)$$

and $J_{ij}$ is the adjoint of the deformation matrix $\partial q_i/\partial a_l$ with

$$\frac{\partial q_i}{\partial a_j}J_{il} = J\delta_{lj}, \quad J_{il} = \frac{\partial J}{\partial(\partial q_i/\partial a_l)}. \quad (2.6)$$

Finally,

$$V_Q(q) = -\frac{\hbar^2}{2m\sqrt{\rho(q)}}\frac{\partial^2\sqrt{\rho(q)}}{\partial q_i \partial q_i} \quad (2.7)$$

is the quantum potential with

$$\rho(q(a,t),t) = J(a,t)^{-1}\rho_0(a). \quad (2.8)$$

The initial data to be appended to the dynamical equations (2.2) and (2.3) is

$$\frac{\partial q_{i0}(a)}{\partial t} = \frac{1}{m}\frac{\partial S_0(a)}{\partial a_k}, \quad \rho_0(a) = |\psi_0(a)|^2. \quad (2.9)$$

Conversely, version (2.1) of the Schrödinger equation may be derived from (2.2) or (2.3) together with the initial data (2.9) [1]. The wavefunction constructed from a solution $q_i$ is given, in polar form, by

$$\psi(x,t) = \sqrt{(J^{-1}\rho_0)|_{a(x,t)}} \exp\left[\frac{i}{\hbar}\left(\int m\partial q_i(a,t)/\partial t|_{a(x,t)} dx_i + f(t)\right)\right] \quad (2.10)$$

(for the determination of the function $f(t)$ see [1]).

A fundamental property of the dynamical equations (2.2) and (2.3) is that the probability is conserved along their solutions:

$$\rho(q,t)d^3q(a,t) = \rho_0(a)d^3a. \quad (2.11)$$

We assert that at time $t$ the congruence of trajectories $q_i(a)$ constitutes the spectrum of possible outcomes of a position measurement, the trajectory density reflecting the quantal probability density. This is easily confirmed by applying the theory to a typical measurement process. As to *which* trajectory is manifested as the outcome of an individual measurement, and how it is connected to corporeal matter, require further interpretative analysis. These problems have in fact been solved but the validity of the trajectory concept of state is not dependent on a particular interpretation of quantum mechanics. We note also that the notion of a trajectory possessing simultaneously well-defined values of position ($q_i$) and momentum ($m\dot{q}_i$) is not in conflict with the uncertainty relations, which comprise correlations in the statistical scatter of sequences of measurement results.

A more elegant approach to the trajectory theory, which brings out several important formal aspects of the approach, is to introduce vector potentials for the wavefunction [2]. These potentials form a set of phase space variables from which the trajectory description is obtained by a canonical transformation. This construction indicates that the connection between the two versions of state is not one-to-one; there is a gauge freedom in the trajectory equations, a relabelling transformation of the orbits, with respect to which the $\psi$ formalism is insensitive.

**3 Representation of the many-body quantum state in terms of functions in 3-space**

A system of $n$ particles with masses $m_r$, $r = 1,\ldots,n$, has an associated wavefunction $\psi(x_1,\ldots,x_n)$ defined in a $3n$-dimensional configuration space where $x_{ri}$, $i = 1,2,3$ represents a set of rectangular Cartesian coordinates. The dynamical equation in this formulation is

$$i\hbar\frac{\partial \psi}{\partial t} = -\sum_{r=1}^{n}\frac{\hbar^2}{2m_r}\frac{\partial^2 \psi}{\partial x_{ri}\partial x_{ri}} + V(x_1,\ldots,x_n)\psi, \quad i=1,2,3, \quad r=1,\ldots,n. \quad (3.1)$$

In a straightforward generalization of the single-body theory of the last section, the $n$-body quantum state may be pictured alternatively as a (single-valued) congruence of curves $q_{ri}(a_1,...,a_n)$ in the 3n-dimensional configuration space where the indices *r,i* collectively range over 3n values, the arguments $a_1,...,a_n$ uniquely label the initial positions $q_{r0i} = a_{ri}$, and the initial density and velocity are specified in accordance with $\psi_0$ [1,2,4]. From the grouping of the indices we see immediately that in this picture each configuration space trajectory is composed of *n* trajectories in three-dimensional physical space, the *r*th trajectory being given by the position vector $q_{ri}$. The whole non-denumerable configuration space congruence is therefore composed of *n* families of trajectories in 3-space. *The n-body quantum state may be represented as a collection of n states in 3-space*.

Note that the trajectories comprising each of the *n* 3-families may cross the same spacetime point, as may trajectories drawn from different 3-families.

As in the one-body case, we can give self-contained first- and second-order renditions of the trajectory version of the many-body wave equation. For the second-order variant the Schrödinger equation becomes a set of *n* Newton-like equations describing the coupled evolution of the set of *n* displacement 3-vectors:

$$m_r \frac{\partial^2 q_{ri}(a_1,...,a_n)}{\partial t^2} = -\frac{\partial}{\partial q_{ri}}\left[V(x_1,...,x_n) + V_Q(x_1,...,x_n)\right]_{x_r = q_r(a_1,...a_n,t)} \quad (3.2)$$

with initial conditions

$$\frac{\partial q_{r0i}(a_1,...,a_n)}{\partial t} = \frac{1}{m_r}\frac{\partial S_0(a_1,...,a_n)}{\partial a_{ri}}, \quad \rho_0(a_1,...,a_n) = |\psi_0(a_1,...,a_n)|^2. \quad (3.3)$$

Here we employ generalizations of the formulas (2.4)-(2.8) obtained by extending the index range. The wavefunction may be constructed from the solutions to (3.2) as follows, an obvious generalization of (2.10):

$$\psi[x_1,...,x_n,t] = \sqrt{\left(J^{-1}\rho_0\right)\big|_{a_r(x_1,...,x_n,t)}} \exp\left[\frac{i}{\hbar}\left(\sum_{r=1}^{n}\int m_r \dot{q}_{ri}(a_1,...,a_n,t)\big|_{a_r(x_1,...,x_n,t)} dx_{ri} + f(t)\right)\right]. \quad (3.4)$$

From (3.2) it is evident that the trajectory $q_{ri}$ is generally coupled with all the other current locations $q_{r'i}$, $r' \neq r$. Hence, if one family of trajectories, say the *r*th, is acted upon by an external force, the whole congruence will generally respond simultaneously to the localized influence. This is how nonlocality is expressed for this notion of state.

Each 3-trajectory is identified by the parameters $a_{ri} = q_{r0i}$. The initial velocity (3.3) of each trajectory generally depends on all the parameters $a_1,...,a_n$ and, as time progresses, each function $q_{ri}$, $r = 1,...,n$, may become dependent on the labels $a_{r'i}$ of the other 3-trajectories $r' \neq r$ due to coupling induced by the dynamical equations. This mutual dependence is how the trajectory model manifests entanglement of the *n*-body wavefunction. We can establish this connection by demonstrating the equivalence of the conditions for independence in the two pictures:

*Theorem of independence*: The wavefunction factorizes into a product of *n* 3-factors,

$$\psi(x_1,...,x_n) = \prod_{r=1}^{n} \psi_r(x_r), \qquad (3.5)$$

if and only if the corresponding 3-space vectors are mutually independent and $\rho_0$ factorizes:

$$q_{ri}(a_1,...,a_n) = q_{ri}(a_r,t), \quad |\psi_0(x_1,...,x_n)|^2 = \prod_{r=1}^{n} |\psi_{r0}(x_r)|^2 \qquad (3.6)$$

*Proof*: Eq. (3.5) implies that the total phase is additive: $S(x_1,...,x_n) = \sum_{r=1}^{n} S_r(x_r)$. From (3.4) this implies that $\dot{q}_{ri} = F_i(q_r)$ whose solution $q_{ri}$ depends just on the parameters $q_{r0i} = a_{ri}$. Hence, the first condition in (3.6) is obeyed and the second condition follows from (3.5). Conversely, $q_{ri} = q_{ri}(a_r)$ implies $\sum_{r=1}^{n} m_r \dot{q}_{ri}(a_1,...,a_n,t)\big|_{a_r(x_1,...,x_n,t)} = \sum_{r=1}^{n} m_r \dot{q}_{ri}(a_r,t)\big|_{a_r(x_r,t)}$ and $J(a_1,...,a_n) = \prod_{r=1}^{n} J_r(a_r)$ so using the second relation in (3.6) we obtain (3.5) from (3.4). □

Applying the conditions of the theorem, (3.2) and (3.3) yield *n* copies of the 1-body formulas.

We can devise hybrid models of the quantum state that mix aspects of both the wavefunction and trajectory approaches by evaluating a selection of arguments of $\psi$ in (3.4) along the trajectories and leaving others as space coordinates. For example, the state may be represented as a space function of just one particle's coordinates, say $x_{ri}$:

$$\phi_r(x_r, a_1,...,a_n) = \psi\left[x_1(a_1,...,a_n),...,x_{r-1}(a_1,...,a_n), x_r, x_{r+1}(a_1,...,a_n),...,x_n(a_1,...,a_n)\right]. \quad (3.7)$$

Inverting, we can derive $\psi$ from a 3-space function: $\psi(x_1,...,x_n,t) = \phi_1(x_1,a_1,...,a_n)\big|_{a_r(x_1,...,x_n,t)}$. In a more symmetrical representation:

$$\psi(x_1,...,x_n,t) = \frac{1}{n}\left[\phi_1(x_1,a_1,...,a_n) + ... + \phi_n(x_n,a_1,...,a_n)\right]_{a_r(x_1,...,x_n,t)}. \qquad (3.8)$$

Thus, *the wavefunction of an n-body system may be derived from the superposition of n 'single-particle' functions in 3-space*. Actually, we can remove all reference to *x* and give an *a*local representation of $\psi$:

$$\Psi(a_1,...,a_n,t) = \psi\left[x_1(a_1,...,a_n,t),...,x_n(a_1,...,a_n,t),t\right]. \qquad (3.9)$$

The hybrid wave equations obeyed by the wavefunction when various sets of $x_{ri}$s are evaluated along trajectories are easily found by transforming the Schrödinger equation appropriately (for the single-body case see [5]).

We have seen that the trajectory construction provides a solution to the old problem of representing a many-body system in terms of a set of 'local' functions, one for each particle [6]. The usual way to associate a state with one particle in a many-body system, say the $r$th, is to employ the partial density matrix obtained by integrating the pure state density matrix over all coordinates except $x_{ri}$. This definition is justified insofar as the partial density matrix accounts for measurements of operators pertaining just to the $r$th particle, and it defines a conserved density. However, the set of $n$ single-body reduced matrices obtained for the whole system does not contain sufficient information to reconstruct the pure state $\psi$ and hence this method is not the basis of an alternative three-dimensional representation of the state. Other methods of three-dimensional representation have been tried, motivated in part by a desire to avoid quantum nonlocality (e.g., [7] and references therein). As we have seen, that quest is unattainable; each 3-state generally depends on the $3n$-3 parameters defining the remainder of the many-body system. Except for the special case treated in the independence theorem above, the configuration space is irreducible and so, whatever formulation is used to represent the quantum state, nonlocality is a generic feature. One cannot remove the irreducible configuration-space dependence of quantum many-body systems by a change of coordinates, as remarked previously [6].

## 4 Identical particles

The use of symmetric or antisymmetric wavefunctions to treat a system of identical particles is generally considered to render the formalism bereft of any means of labelling or distinguishing the individual particles. In contrast, in the trajectory formulation the $r$th particle in a system of $n$ identical particles is distinguished by the temporal continuity of the $r$th family of orbits $q_{ri}(t)$. Identity is expressed through symmetry properties of the trajectories: under exchange of the $r$th and $r'$th labels the many-body state obeys the relation

$$q_{ri}(a_1,...,a_{r'},...,a_r,...,a_N) = q_{r'i}(a_1,...,a_r,...,a_{r'},...,a_N) \qquad (4.1)$$

together with symmetrization of the initial conditions (3.6). The symmetry constraints are reflected in the spatial characteristics of the paths generated by inter-trajectory forces. According to this model 'identical' and 'distinguishable' are compatible notions.

## 5 Conclusion

The trajectories we have introduced are structures that may be discerned in the wave field, that is, lines of probability flow. But they are not just aspects of the field; the trajectories provide, together with the appropriate initial conditions on their density and velocity, an alternative conception of the state of the system so that the wave amplitude may be dispensed with and regarded as a derived quantity. The two pictures provide mutually illuminating ways of seeing a single system, each bringing out aspects not present, or only hinted at, in the other. For example, the second-order trajectory dynamical equation makes explicit reference to force as the propeller of quantum propagation, a notion that features only indirectly in the wavefunction approach. Here we also observed that, in the alternative description, the state of an $n$-body system is a set of $n$ families of interlacing spacetime trajectories and that such a notion is compatible with the quantum concepts of entanglement and identity.

[1] P. Holland, *Ann. Phys. (NY)* 315, 505 (2005)